\index{\cite{\footnote{\pageref{\ref{•}}}}}

\documentclass[iop]{emulateapj-rtx4}

\usepackage{psfig}
\usepackage{epsf}
\usepackage{graphics}
\usepackage{color}
\usepackage{lscape}
\usepackage{graphicx}
\usepackage{verbatim}
\usepackage{graphicx}

\usepackage{morefloats}

\newcommand{\HII}{H\,{\sc ii}}
\newcommand{\arcs}{\arcsec}

\newcommand{\Spitzer}{{\it Spitzer }}

\shorttitle{WISE Nearby Galaxies -- I}
\shortauthors{Jarrett et al. 2012a}

\begin{document}

\title{Constructing a  WISE High Resolution Galaxy Atlas\\
}

\author{T.H. Jarrett\altaffilmark{1},  F. Masci\altaffilmark{1},   C.W. Tsai\altaffilmark{1},
S. Petty\altaffilmark{2}, M. Cluver\altaffilmark{3}, Roberto J. Assef\altaffilmark{4,11}, D. Benford\altaffilmark{5}, A. Blain\altaffilmark{6}, C. Bridge\altaffilmark{1},
 E. Donoso\altaffilmark{1}, P. Eisenhardt\altaffilmark{4}, J. Fowler\altaffilmark{1},
 B. Koribalski\altaffilmark{7},
 S. Lake\altaffilmark{2},  James D. Neill\altaffilmark{1}, M. Seibert\altaffilmark{8}, K. Sheth\altaffilmark{9},
 S.A. Stanford\altaffilmark{10}, E. Wright\altaffilmark{2}
 }

\altaffiltext{1}{California Institute of Technology, Pasadena, CA 91125, USA}
\altaffiltext{2}{UCLA}
\altaffiltext{3}{Anglo-Australian Observatory}
\altaffiltext{4}{Jet Propulsion Laboratory}
\altaffiltext{5}{GSFC}
\altaffiltext{6}{Leicester University}
\altaffiltext{7}{Australian National Telescope Facility}
\altaffiltext{8}{Carnegie Institute}
\altaffiltext{9}{NRAO}
\altaffiltext{10}{UC Davis}
\altaffiltext{11}{NASA Postdoctoral Program Fellow}


\begin{abstract}

After eight months of continuous observations,
the Wide-field Infrared Survey Explorer (WISE)
mapped the entire sky at 3.4 $\mu$m, 4.6 $\mu$m, 12 $\mu$m and 22 $\mu$m.
We have begun a dedicated WISE High Resolution Galaxy Atlas (WHRGA) project
to fully characterize large, nearby galaxies and produce a legacy image atlas
and source catalogue.
Here we summarize the deconvolution technique used to significantly
improve the spatial resolution of WISE imaging, specifically designed to
study the internal anatomy of nearby galaxies.
As a case study, we present results for
the galaxy NGC\,1566, comparing the WISE super-resolution image processing to that
of \Spitzer, GALEX and ground-based imaging.  The is the first paper in a two
part series;  results for a much larger sample
of nearby galaxies is presented in the second paper.

\end{abstract}


\pagebreak

\section{Introduction}

For nearly three decades now, starting with IRAS in the early 1980's and continuing today with the \Spitzer and the AKARI Space Telescopes, the infrared properties
of galaxies have been explored at ever increasing sensitivity, spatial and spectral resolution.   The \Spitzer Infrared Nearby Galaxies Survey
(SINGS; Kennicutt et al. 2003) represents the `gold standard' study of nearby galaxies, employing every infrared instrument of \Spitzer to study in detail the properties
of 75 nearby `representative' galaxies.  A larger \Spitzer imaging sample is found in the SINGS follow-up project, Local Volume Legacy (LVL).
Expanding the sample to several thousand galaxies, the \Spitzer Survey of Stellar Structure in Galaxies (S4G; Sheth et al. 2010)
continued the SINGS and LVL surveys through the two short (near-infrared) wavelength bands of IRAC (3.6 and 4.5 $\mu$m), focusing on the internal stellar
structure of galaxies.

Following closely in succession to the AKARI all-sky survey (Murakami et al. 2006),
the latest generation infrared space telescope, the Wide-field Infrared Survey Explorer (WISE),  expands these powerful surveys through its
all-sky coverage and mega-pixel cameras,
capable of constructing large, diverse
and complete statistical samples--for both the near-infrared and mid-infrared windows--sensitive to both stellar structure (as with S4G) and
interstellar processes (as with SINGS).
WISE was specifically designed and optimized to detect and extract point source information.
Detection, for example, was carried out using co-addition of image frames that were constructed
with a resampling method based on
a matched filter derived from the WISE point spread function (PSF).  As a consequence, this interpolation method tends to smear the images,
making them less optimal for detection and characterization of resolved sources.
However, owing to the stable PSF for all four WISE bands, there is a way to apply deconvolution
techniques to recover from the smearing and further improve the spatial resolution.
In this first paper of a two part series, we  demonstrate how the angular resolution of WISE may be enhanced to achieve information on
physical scales
comparable to those of \Spitzer imaging, that enable detailed study of the internal anatomy of galaxies.
We employ a resolution enhancement technique known as the Maximum Correlation
Method (MCM;  Masci \& Fowler 2009) to construct the WISE High Resolution Galaxy Atlas (WHRGA), consisting of several thousand
nearby galaxies.

The WHRGA will comprise a complete mid-infrared source catalog and high-resolution image atlas
of the largest angular-sized galaxies in the local universe.
In this first paper we summarize the MCM algorithm and demonstrate its performance
using WISE, \Spitzer and GALEX imaging of nearby spiral galaxy NGC\,1566.
In the second paper (Jarrett et al. 2012b;  hereafter, referred to
as Paper II), we demonstrate the early results
of the WHRGA-project for a sample of 17 galaxies,  all observed by \Spitzer and GALEX, chosen to
be of large angular size, diverse morphology, and covering a range in color, stellar mass
and star formation.   In addition to basic photometry, source characterization and
surface brightness decomposition, Paper II also derives
star formation rates and stellar masses.

This first paper is organized as follows.    Section 2 provides more technical information about
the WISE mission and data products.  Section 3 the ancillary data used to compare to WISE, including \Spitzer and
GALEX imaging.
Section 4 we outline the MCM
 image deconvolution- method and illustrate its performance using a set of simulated observations.
Section 5 we focus on a case study of NGC\,1566 to demonstrate the super-resolution performance using real WISE imaging
data.
All reported magnitudes are in the Vega System (unless otherwise specified).

\section{WISE Mission and Data Products}

The NASA-funded Medium-Class Explorer mission, WISE, consists of a 40-cm space infrared telescope, whose science instrumentation
includes 1024x1024 pixel Si:As and HgCdTe arrays, cooled with a two-stage solid hydrogen cryostat.
Dichroic beamsplitters allow simultaneous images in four mid-infrared bands, each covering a $47' \times 47'$ field of view.
The duty cycle was 11\,s,
achieved using a
 scan mirror that stabilizes the line-of-sight while the spacecraft scans the sky, achieving an angular resolution of
  $\sim$6$\arcsec$ in the short bandpasses and $\sim$12$\arcsec$  in the longest bandpass.
 Multiple, overlapping frames are combined to form deeper coadded images.
Launched in December of 2009 into a sun-synchronous polar orbit, over a time span of eight months WISE
completed its primary mission to survey the  entire sky in the  3.4, 4.6, 12 and 22 $\mu$m infrared bands
with 5\,$\sigma$ point-source sensitivities of at least 0.08, 0.11, 0.8, and 4 mJy, respectively
(Wright et al. 2010) and considerably deeper sensitivities at higher ecliptic latitudes (Jarrett et al. 2011).

\smallskip

Detailed in the WISE Explanatory Supplement (Cutri et al. 2012)\footnote{
http://wise2.ipac.caltech.edu/docs/release/},
``Atlas" Images. \footnote{Public release WISE co-added images are referred to as ``Atlas" images, available through
http://irsa.ipac.caltech.edu/applications/wise/}
are created from single-exposure frames that touch a pre-defined
1.56$^\circ\times$1.56$^\circ$ footprint on the sky.
For each band, a spatially registered image is produced by interpolating and coadding multiple 7.7/8.8\,s single-exposure images
onto the image footprint.
To suppress copious cosmic rays and other transient events that populate the single-exposure frames, time-variant pixel outlier
rejection is used during the co-addition process. The resulting sky intensity ``Atlas" mosaics are 4095$\times$4095 pixels with 1.375$\arcsec$/pixel scale,
providing a 1.56$^\circ\times$1.56$^\circ$ wide-field.
In addition to the sky intensity mosaics, 1\,$\sigma$ uncertainty maps (tracking the error in intensity values) and
depth-of-coverage maps
are part of the standard products.
The number of frames that are coadded depends on the field location relative
to the ecliptic: those near the equator will have the lowest coverage (typically 12 to 14 frames), while those near the poles have the
highest coverage ($\gg$1000 frames).

The WISE All-Sky public data release in March 2012 includes imaging and source catalogs,  available
through the Infrared Science Archive (IRSA).
It should be emphasized that the WISE Source Catalog
is designed, optimized and calibrated for point sources.
The complexity of detecting and measuring resolved
sources was beyond the resources of the WISE Science Data Center (WSDC) processing.  As a consequence, the WISE archive
and public release catalogs have either completely missed nearby galaxies or, even worse, their integrated fluxes are systematically
underestimated (because they are measured as point sources) and often chopped into smaller pieces.
However, the WISE public-release
imaging products do capture
resolved and complex objects.  One of the goals of this current study is to use new image products to characterize and assess the
quality of source extraction for resolved galaxies observed by WISE.  We apply image resolution-enhancement techniques and compare
the resulting measurements with those extracted using \Spitzer imaging (Paper II).

\begin{figure*}[ht!]
\begin{center}
\includegraphics[width=17cm]{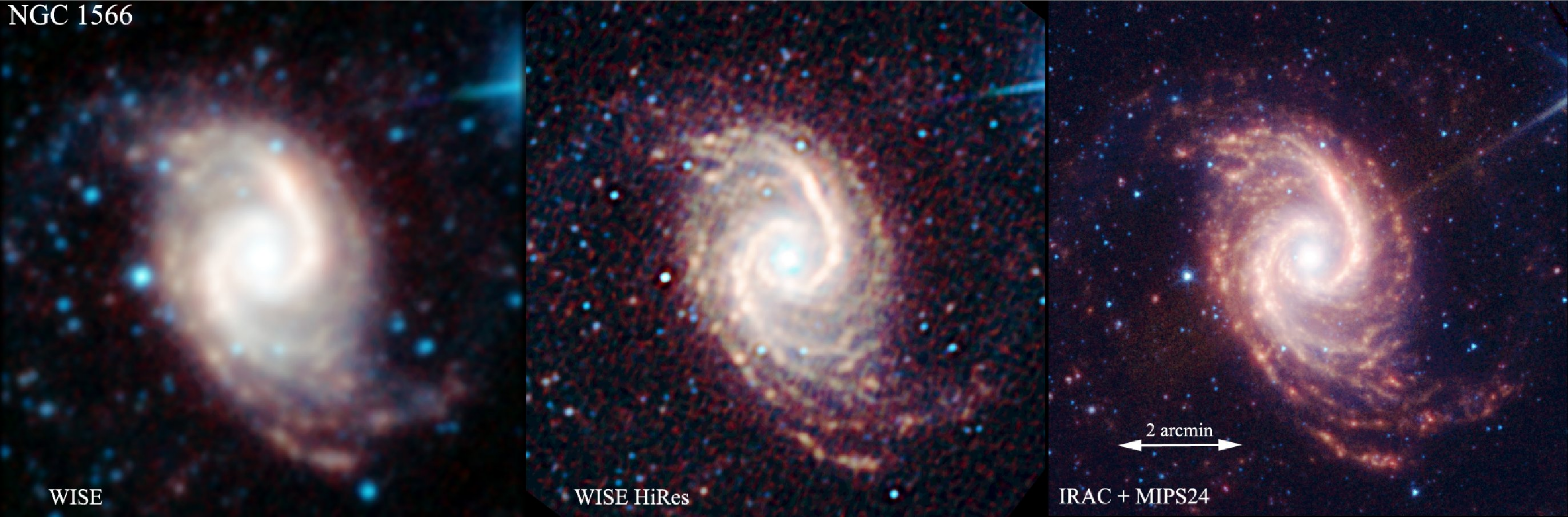}
\caption[HiRes2]
{\small{WISE and {\it Spitzer} composite of NGC\,1566.
 The left panel
shows the nominal (public-release) WISE mosaic, where the
colors correspond to WISE bands:  3.4 $\mu$m (blue),
4.6 $\mu$m (green),
12.0 $\mu$m (orange),
22 $\mu$m (red).
The middle
panel the
MCM-HiRes
spatial resolution enhancement of the WISE mosaics.
The right panel shows IRAC+MIPS mosaic, where
the colors correspond to:  3.6 $\mu$m (blue),
4.5 $\mu$m (green),
5.8 $\mu$m (yellow),
8.0 $\mu$m (orange),
24 $\mu$m (red).
}}
\label{HiRes2}
\end{center}
\end{figure*}

\section{Ancillary Data}

\subsection{\Spitzer SINGS Imaging}

In this first paper we focus on the spiral galaxy NGC\,1566
\footnote{It has ecliptic coordinates of 32.07, -73.35 degrees, which means
it has very good coverage with WISE.}, which
was part of the \Spitzer-SINGS survey of nearby galaxies. It is located
at a distance of 9.5 Mpc (based on four separate Tully-Fisher measurements;
Willick et al. 1997; Tully 1988), has a morphological class type of SAB(rs)bc
and includes a Seyfert-1 nucleus (NASA Extragalactic Database).
The SINGS team has
provided enhanced-quality spectroscopy and imaging mosaics that is
available to the publich through the \Spitzer Data Archives\footnote
{http://data.spitzer.caltech.edu/popular/sings/20070410\_enhanced\_v1/}.
For this work, we utilize \Spitzer-IRAC and \Spitzer-MIPS24 imaging, which
are fully calibrated with astrometry and photometric solutions.
Additionally, the SINGS team have provided ancillary ground-based imaging,
of which we use the optical B-band and H$\alpha$ data products.

\subsection{GALEX Imaging}

GALEX FUV (0.1516 $\mu$m) and NUV (0.2267 $\mu$m) images of NGC\,1566 were obtained from the GALEX Medium
Imaging Survey (MIS; Martin et al. 2005), which were
processed using the standard
GALEX pipeline (Morrissey et al. 2005, 2007).
The MIS reaches a
limiting NUV magnitude of
23 (AB mag)
through multiple eclipse exposures that are
typically 1 ks or greater in
duration, while azimuthal averaging reaches surface brightness depths of $\sim$30 to 31 mag\,arcsec$^{-2}$
(AB mag).


\section{WISE High-Resolution Reconstruction}

The nominal spatial resolution of WISE, $\sim$6$\arcsec$ in the three short-wavelength bands and 12$\arcsec$ in the 22$\mu$m band,
is relatively poor compared to ground-based and space-based infrared observations; e.g., it is 3$\times$ larger than that of {\it Spitzer} IRAC imaging.
For nearby galaxies, the internal structures are smeared and strongly blended from the nucleus to the disk boundary; consequently, it is a challenge to
decompose the internal anatomy or
make a detailed comparison with ground-based imaging (e.g., H\,$\alpha$ line maps) and
S4G IRAC imaging.
Except for the few largest galaxies, only the global properties are easily obtained from nominal WISE imaging.  We can, however, recover information that is lost within the WISE primary beam using deconvolution methods.  These take full advantage of the
relatively stable and well-characterized point spread function (PSF) of  WISE.   The first method is the widely used
Variable-Pixel Linear Reconstruction, or ``drizzle" technique of co-addition.  The second is a true deconvolution technique detailed below.

The WSDC has developed a generic co-addition and resolution enhancement
(HiRes) tool specifically designed to operate on WISE single-exposure image frames. This tool
produces science-quality image products with statistically-validated uncertainty
estimates on fluxes. The HiRes algorithm is based on the Maximum Correlation
Method (MCM) of Masci \& Fowler (2009) and is an extension of the classic
Richardson-Lucy deconvolution algorithm, originally implemented to boost the scientific return from IRAS approximately 20
years ago (Aumann et al. 1990; Fowler \& Aumann 1994; Cao et al. 1997), and is still provided as an online
service to users\footnote
{http://irsa.ipac.caltech.edu/IRASdocs/hires\_over.html}.

The scientific purpose of MCM-HiRes is to significantly enhance the spatial resolution of images while also
conserving the integrated flux and maintaining photometric integrity with the extended, low surface brightness emission.
Fig. \ref{HiRes2} qualitatively demonstrates both requirements; after resolution enhancement the WISE imaging of the
galaxy NGC\,1566 is greatly improved (panel 1 vs. 2) and resembles the \Spitzer IRAC+MIPS-24 composite of the galaxy (panel 3).
The spiral arms, filaments and lower surface brightness features in the WISE HiRes image are also seen
in the \Spitzer composite, there are no obvious artifacts or isolated features that are unique to the HiRes image relative to {\it Spitzer}.
The images are shown with four colors, where each color is assigned to
a WISE band:  blue $\leftrightarrow$ W1 (3.4 $\mu$m), cyan $\leftrightarrow$ W2 (4.6 $\mu$m),  orange $\leftrightarrow$ W3 (12 $\mu$m)
and red $\leftrightarrow$ W4 (22 $\mu$m).   Stellar light from the old, evolved population will appear blue/green
and tends to concentrate in the nucleus and bulge regions.   The ISM, warmed and excited by star formation, will
appear yellow/orange, delineating \HII\ and photo-dissociation regions (PDRs) as well as warm dust emission (red) from
the disk.   Later (Section XX) we conduct a quantitative comparison between WISE, \Spitzer and GALEX.

\begin{figure*}[ht!]
\begin{center}
\includegraphics[width=15cm]{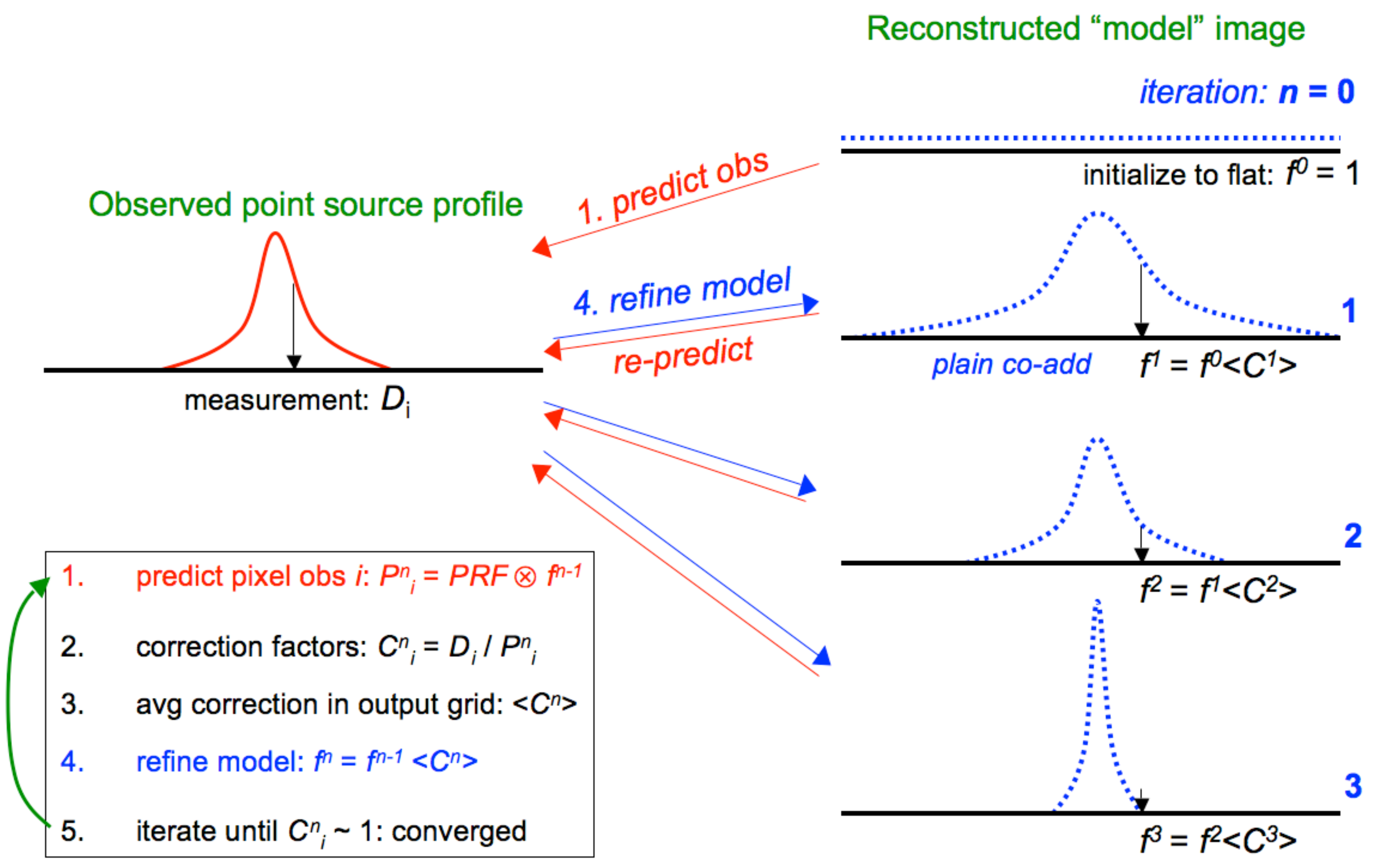}
\caption[MCM]
{\small{Illustrating the iterative deconvolution process of the
Maximum Correlation
Method,  an extension of the classic
Richardson-Lucy iterative algorithm.
At each iteration, the model is refined and improved using
spatially dependent correction factors C such that the ``convolved model" reproduces the measurements within the noise,
equivalent to C converging to unity.
}}
\label{MCM}
\end{center}
\end{figure*}

\smallskip

The overall goal of MCM is to yield a
``model" of the sky that is consistent with the observations to within
measurement error; see Fig \ref{MCM}. The baseline algorithm assumes no prior information or
regularizing constraints like in previous approaches, although use of prior
(e.g., cross-wavelength) information is optional. MCM allows for
non-isoplanatic (spatially-varying) point spread functions, noise-variance
weighting, {\it a-posteriori} uncertainty estimation,
ringing-suppression\footnote{Like most deconvolution methods, MCM can lead to ringing artifacts in the reconstructed image. This
limits super-resolution, i.e., when attempting to go well beyond the diffraction limit of an
imaging system.},
statistically motivated convergence criteria and metrics to assess the
quality of HiRes solutions, and use of redundant overlapping exposures
to optimize the signal-to-noise. The tool also includes preparatory steps
such as background-level and photometric-throughput matching, and
outlier/bad-pixel detection and masking.    To follow, we validate its performance
with simulations and then with real imaging from WISE that is compared
with {\it Spitzer}, GALEX and ground-based imaging.

\begin{figure*}[h!]
\begin{center}
\includegraphics[width=16cm]{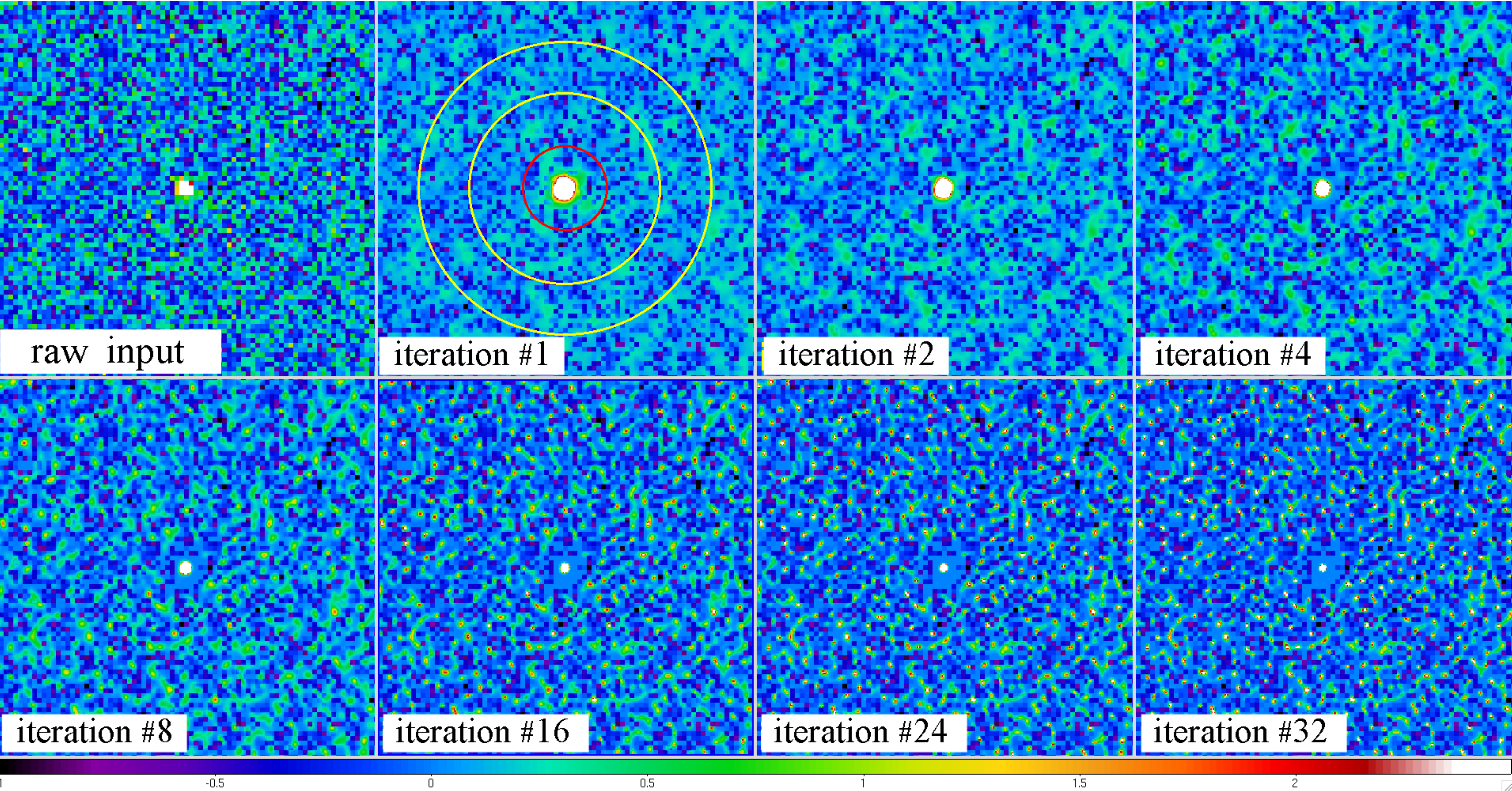}
\caption[SIM1]
{\small{Evolution of the pixel noise distribution using the MCM HiRes-MCM technique.
A series of intensity images containing
simulated Poission noise and a single point-source in the middle.
The first is the raw
input simulated image, followed by the outputs after seven
different MCM iterations. These illustrate
the evolution of the noise structure. The ``iteration 1'' output image
shows an overlay of the source aperture and background annulus used for
the photometric analysis 
.
}}
\label{SIM1}
\end{center}
\end{figure*}

\subsection{Simulations}

We have explored the impact of the MCM  process
on photometric flux and noise measurements extracted from resolution-enhanced images
using simulations. A simulation here is useful for two reasons:
First, it enables us to validate the accuracy of outputs given
knowledge of the truth, and second, it provides us with a method
for unambiguously computing the output noise (in response to the
input) by simulating repeated noise realizations sampled from a
known input noise-distribution model. The simulation uses a single
input image with input/output pixel sizes similar to that used in
processing of the WISE images in this work and Paper II. We assumed a
spatially flat background of 30 DN and added a point-source of total
flux 3000 DN at the center of the image, convolved with the W1 native PSF.
We could have used another WISE band, but any one band suffices to
illustrate the HiRes performance in general. We then added
Poisson noise to this image with variance $\sigma^2 = DN/g$, where
we assumed an electronic gain factor of $g=1$ for simplicity.
Different input noise models and/or distributions don't change our
general conclusions. 500 independent noise realizations (or trials) were
simulated for the input image, and each trial was processed through
the MCM-HiRes algorithm to seven different iterations: 1, 2, 4, 8,
16, 24, and 32. Output HiRes images at these iterations for a
{\it single} simulation-trial are shown in Figure \ref{SIM1}. At the
top-left of this figure is the input raw simulated image for this
trial. We used the same MCM processing parameters as used for
the WISE image reconstructions in this work and Paper II.

\begin{figure*}[h!]
\begin{center}
\includegraphics[width=16cm]{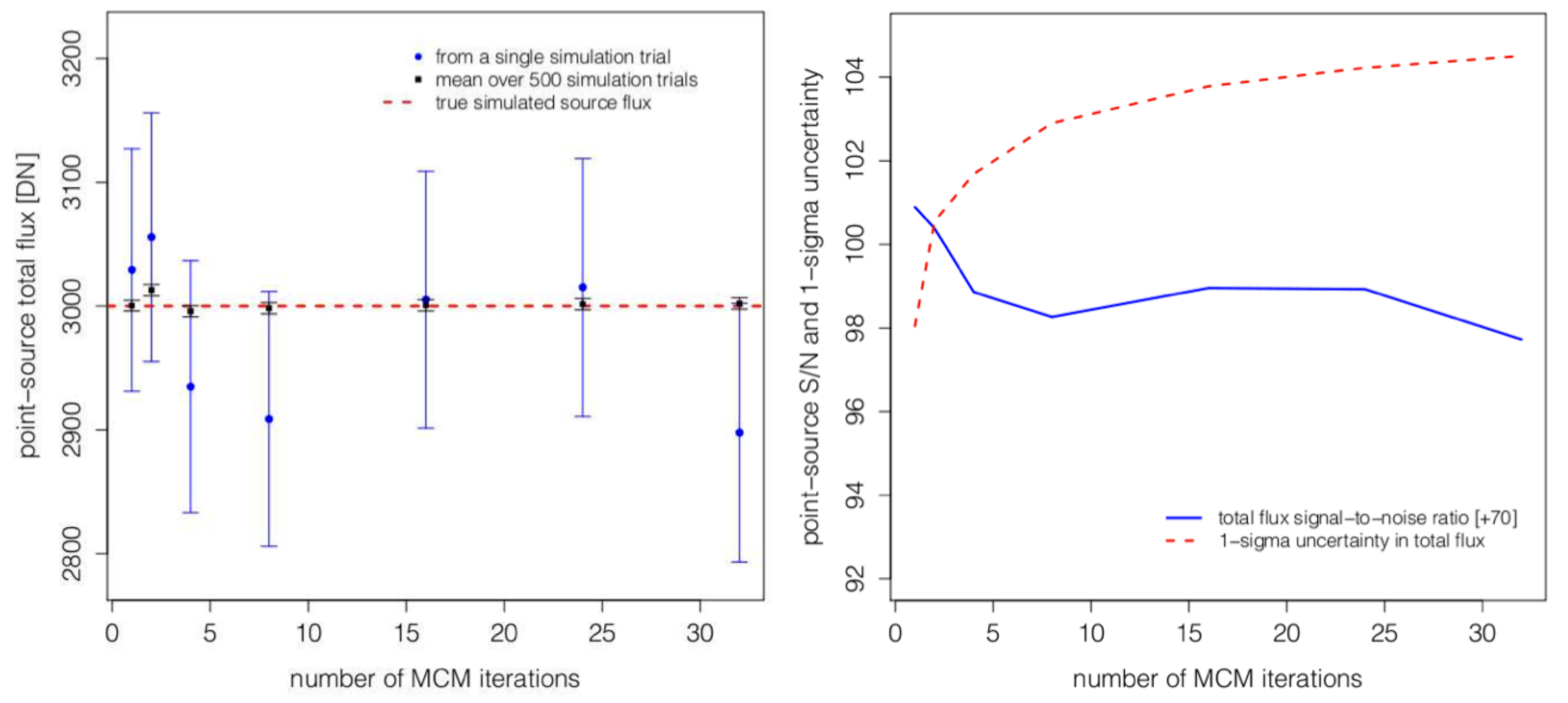}
\caption[SIM2]
{\small{Left:\,
Total flux of the central point-source shown in Figure \ref{SIM1} derived from
aperture photometry as a function of iteration number. Measurements were
made on image outputs from single simulation trials ({\it blue circles}) as
well as an averaged stack of 500 trials at each iteration ({\it black
squares}). The {\it red dashed line} shows the true input simulated
flux (= 3000 DN). These results show that integrated flux measurements
made on the HiRes'd images are consistent with the truth (within measurement
errors) and hence unbiased.
Right:\, Evolution of the Signal-to-Noise (S/N) ratio with MCM iteration
for the single simulation-trials ({\it blue curve}).  An offset of 70 was added for display purposes.
Also shown is the 1-$\sigma$ uncertainty used for the S/N estimates
({\it red dashed curve}). This represents the standard-deviation in the
measured point-source flux (using aperture photometry) over 500
independent simulation trials.
}}
\label{SIM2}
\end{center}
\end{figure*}

\begin{figure*}[ht!]
\begin{center}
\includegraphics[width=16cm]{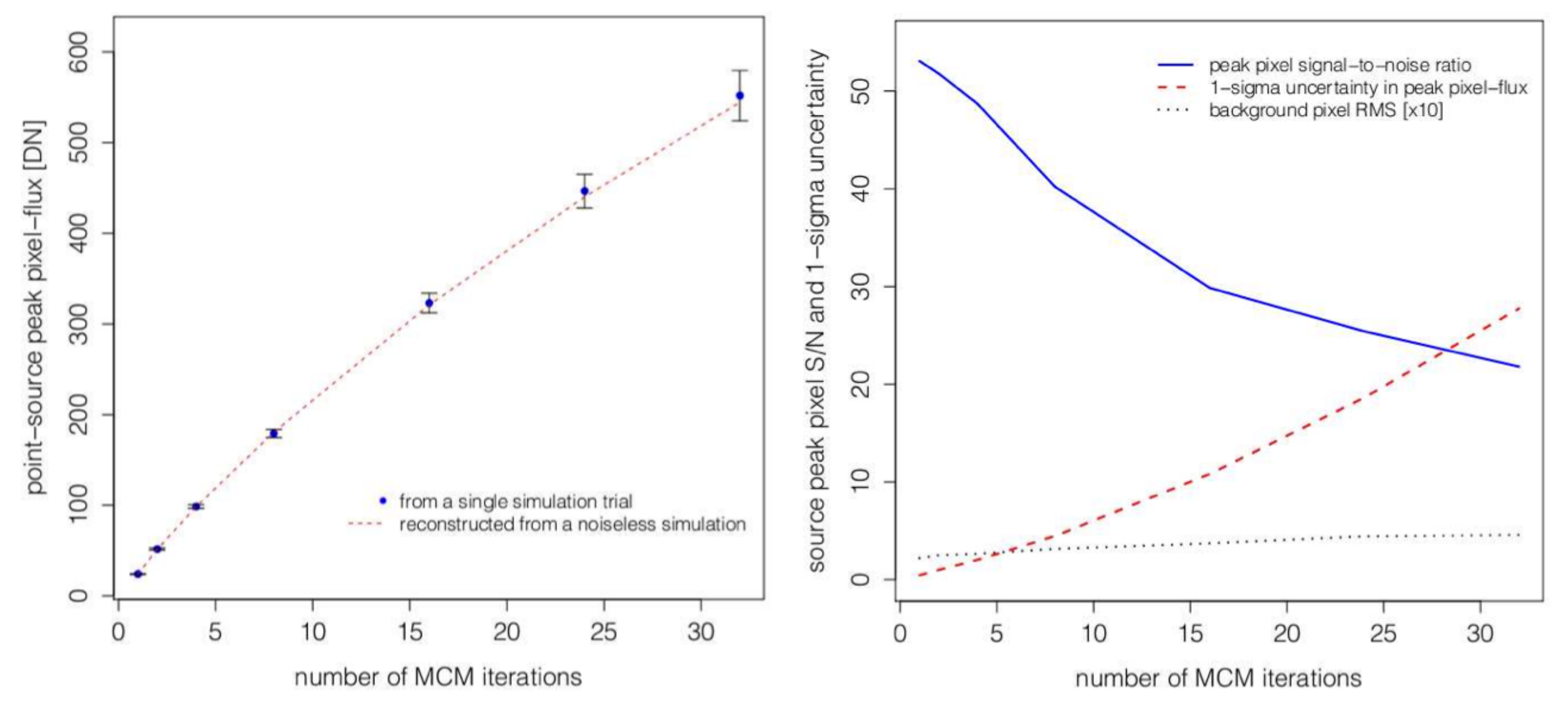}
\caption[SIM3]
{\small{Left:\,
Flux value in the peak pixel of the point-source shown in
Figure \ref{SIM1}  as a function of iteration number. Measurements were made on
image outputs from single simulation trials ({\it blue circles}). The
{\it red dashed line} shows the true reconstructed peak pixel flux one
would obtain if noise were absent. In general, this plot shows the
rate at which flux from the wings of a point-source is forced into
its peak with increasing MCM iteration number.
Right:\, Evolution of the peak-pixel S/N ratio with MCM iteration
number for the single simulation-trial measurements.
The 1-$\sigma$ uncertainty used for these S/N
estimates is shown by the ({\it red dashed curve}). This represents
the standard-deviation in the measured {\it peak-pixel} source flux over
500 independent simulation trials. Also shown is the RMS in background
noise fluctuations (per pixel) versus iteration number ({\it black
dotted curve}). The latter were multiplied by 10 for clarity.
}}
\label{SIM3}
\end{center}
\end{figure*}

One can see in Figure \ref{SIM1} that at low iterations, the noise becomes
spatially correlated, then with increasing iteration, the positive noise spikes and the
point-source at the center gradually ``decorrelate'' and sharpen.
Here the point-source FWHM went from
$\simeq 5.8\arcsec$ (raw input simulated image) to $\simeq 1.75\arcsec$
(iteration 32), equivalent to an increase of 11$\times$ in the flux per
solid angle at the point source position.

Figure \ref{SIM2}-left shows the dependence of the measured point-source flux
(using standard aperture photometry) on the number of MCM iterations
from the single simulation-trial images and the average over all 500
independent trials at each iteration number. This shows that
the measured fluxes are consistent with the true flux (within
measurement error) or what one would measure in an ``observed'' noisy
image. This is an important requirement for any deconvolution process.
Figure \ref{SIM2}-right shows the behavior of the photometric Signal-to-Noise (S/N)
ratio and 1-$\sigma$ uncertainty in the integrated point-source flux
(i.e., the single simulation-trial error bars in Figure \ref{SIM2}-left) as
estimated from the standard-deviation of measurements over 500
independent trials. In general, the S/N
is expected to smoothly decrease monotonically with
iteration number. The hump in S/N at iterations 16 and 24 is due to
noise in the single-trial flux measurements at these
iterations. The decrease in the integrated
S/N is a consequence of noise amplification in general with
increasing iteration number (red dashed curve in Figure \ref{SIM2}-right).
Noise amplification is most prevalent when the "ringing suppression" option is used in
MCM (adopted throughout this work). It leads to an asymmetric noise distribution
with positive noise tail that grows progressively with increasing iteration
(for details, see Masci \& Fowler 2009).
However, the observed drop in integrated S/N is relatively small, and is
$\approx 6$\% over 20 MCM iterations -- the maximum number
of iterations used in this work.

Similarly, figure \ref{SIM3} illustrates the evolution of the peak pixel-flux of our
simulated point-source, including its 1-$\sigma$ uncertainty and
signal-to-noise ratio as a function of MCM iteration number.
All measurements pertain to single simulation trials. It
shows the effective rate at which the flux of a point-source is
forced into its peak with increasing iteration (assuming the
WISE pixel sampling). The increase in peak pixel-flux is a factor
of $\approx 23$ from iteration 1 to 32. Compared to the uncertainties
in integrated flux measurements (Figure \ref{SIM2}), the relative increase
in the peak-flux uncertainty with iteration number (Figure \ref{SIM3}) is
appreciably greater. This implies local (pixel-scale) noise-fluctuations
are more prone to amplification by the MCM process
especially when the ringing suppression option is utilized (see above). Furthermore, we find
the amount of amplification depends on the input noise level. For example,
Figure \ref{SIM3} also shows the evolution of the background pixel spatial-RMS
(dotted curve, $\times$10 for clarity). The input background (Poisson)
noise simulated here is a factor of 8$\times$ (in $\sigma$) below that at
the point-source peak position, and the background RMS varies by only
a factor of $\approx 2$ from iteration 1 to 32. Meanwhile, the peak
pixel 1-$\sigma$ flux uncertainty varies by a factor of $\simeq 60$
over the same iteration interval. This shows that MCM is inherently
a non-linear process in the reconstruction of signals in the presence
of noise.

To summarize, MCM-HiRes is
is capable of generating science quality image
products with uncertainty estimates on fluxes. Testing on \Spitzer and WISE
imaging has shown that it can achieve a factor of $\sim$3 to 4 increase in
resolution per axis (using 10 to 20 iterations), corresponding to at least an order magnitude increase
in flux per unit solid angle (Fig 1.; see also Paper II). The gain in resolution improves with increasing
survey depth-of-coverage since multiple frame overlaps will provide better
sampling of the PSF.  Nonetheless, the PSF at all WISE bands is sampled at
slightly better than the Nyquist rate so that optimal enhancement will
be possible at even the lowest depths-of-coverage.
Next, we look in more detail at the HiRes performance for NGC\,1566.

\section{HiRes Performance:  case study of NGC\,1566}

The nearby spiral galaxy NGC\,1566 (D = 9.5 Mpc) is used to demonstrate the resolution
gain using the MCM-HiRes deconvolution technique (Section 3.3).  In comparison to
\Spitzer imaging,  the WISE HiRes reconstructed images are, in a qualitative sense, remarkably similar,
as shown in Fig. \ref{HiRes2}.  Note that NGC\,1566 was well covered by WISE,
due to its location at high ecliptic latitude, and thus we expect even better performance with this
galaxy;  the performance  gain from a more diverse set of galaxies is presented in Paper II.
Here we consider narrower point source profiles in
the field of the NGC\,1566 galaxy, the improved spatial resolution of the galaxy in comparison to Atlas and `drizzle' co-addition
images of WISE, and the performance with extended emission and lower surface brightness features.

\smallskip

\begin{figure*}
\begin{center}
\includegraphics[width=17cm]{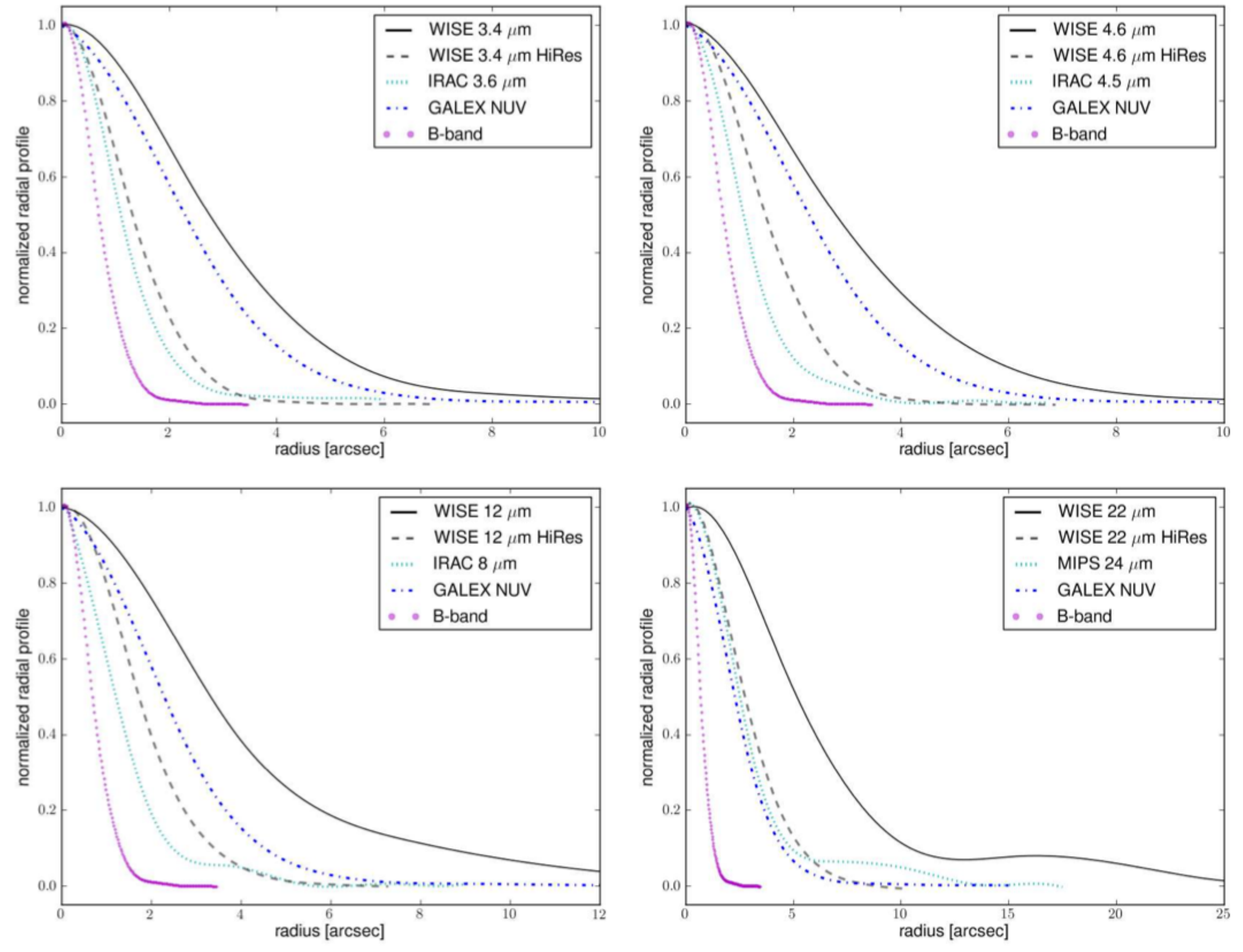}
\caption[Rad]
{\small{Azimuthally-averaged radial point source profiles in the NGC 1566 field, comparing WISE with IRAC, GALEX and
optical ground-based imaging.  The panels are separated by the WISE
bands: 3.4, 4.6, 12 and 22 $\mu$m.  The solid black line depicts the
`drizzle' profile, and the dashed grey line the HiRes profile.
}}
\label{Rad}
\end{center}
\end{figure*}

The improvement in the spatial resolution, as traced
by foreground stars located near NGC\,1566, is shown in Fig. \ref{Rad}. The WISE `drizzled' image
results are shown with the black solid line, and the dashed black line the MCM-HiRes results for a point source.
The FWHM for the `drizzle' images is   5.9, 6.5, 7.0 and 12.4 arcsec, for W1, W2, W3 and W4 respectively,
compared to 2.6, 3.0, 3.5 and 5.5 arcsec, for W1, W2, W3 and W4 respectively for HiRes.  The point source
FWHM results are summarized in Table 1.
At the distance of 9.5 Mpc,
the HiRes is sampling physical scales down to 120 pc.    The figure also shows the PSF beams for
drizzled \Spitzer IRAC and MIPS-24 imaging, and the GALEX NUV and B-band imaging.  In the case of IRAC, the FWHM is
2.1, 2.2, 2.3 arsec for IRAC-1, IRAC-2 and IRAC4, respectively.   The MIPS-24 beam FWHM  is 5.2 arcsec.
The GALEX NUV has an effective beam FWHM of 4.8 arcsec, while the optical B-band image has relatively
good seeing, FWHM = 1.3 arcsec.    It is clear that improvement in spatial resolution is significant for the WISE imaging,
achieving scales that are similar to those of \Spitzer imaging, and at the shorter wavelengths better than those of GALEX.
Next we inspect the performance on the galaxy itself.

\begin{figure*}[ht!]
\begin{center}
\includegraphics[width=17cm]{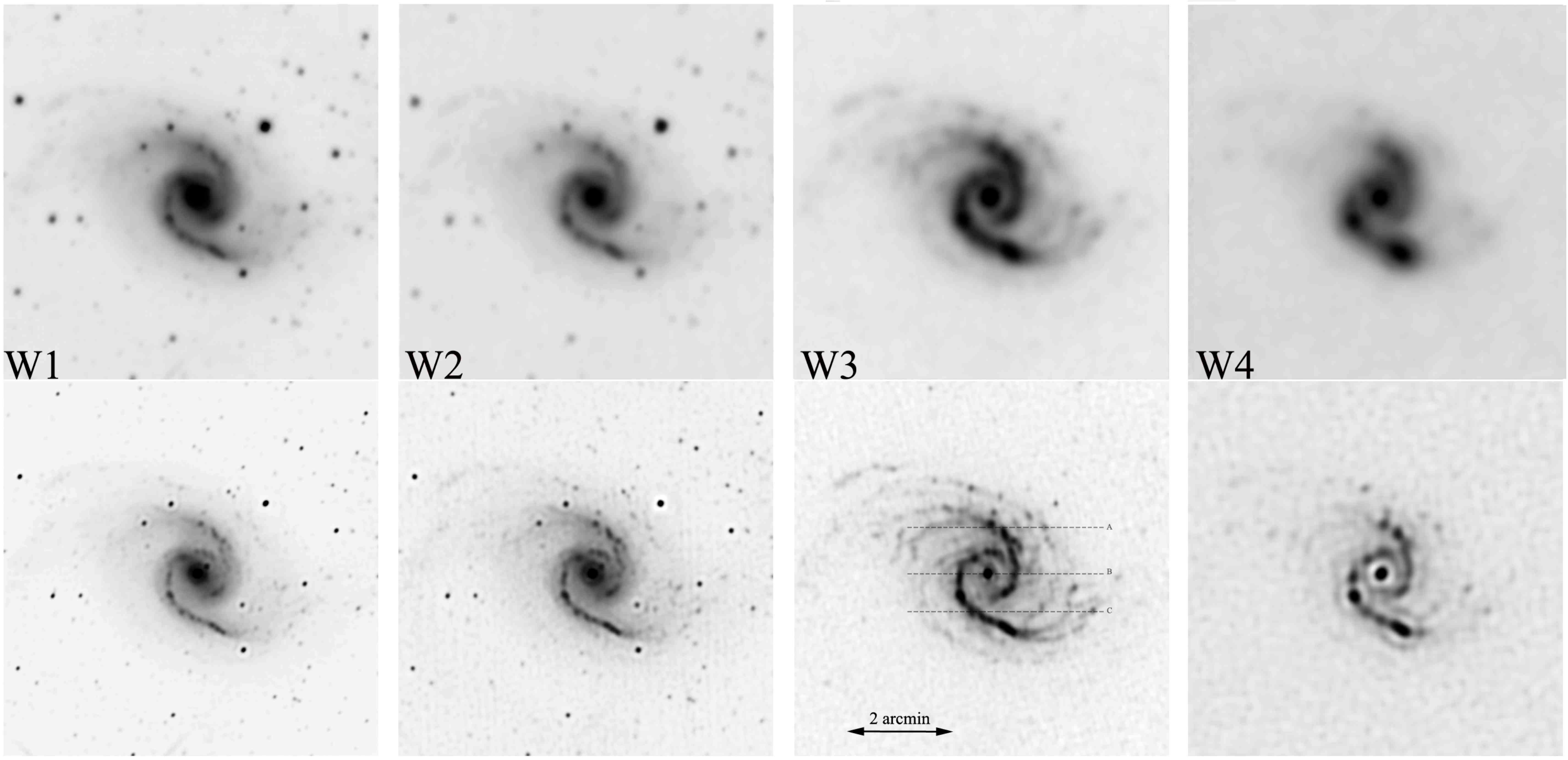}
\caption[HiRes]
{\small{WISE view of NGC\,1566.  The top panels show the
3.4 $\mu$m,
4.6 $\mu$m,
12.0 $\mu$m and
22 $\mu$m channels with nominal WISE mosaic construction.
The bottom panels show the same mosaics after MCM
spatial resolution enhancement.   The dashed lines (A, B and C) shown in the
W3 panel denote the location of the line profile comparisons between
nominal and high-resolution presented in Fig. \ref{line cuts}.
}}
\label{HiResBW}
\end{center}
\end{figure*}

Fig 1. shows a color composite of all WISE bands simultaneously, comparing the nominal Atlas
to HiRes performance.  Decomposing into the individual components,
Fig. \ref{HiResBW}, the HiRes improvement (lower panels) is clearly evident across all bands:
the spiral arms, disk star-formation aggregates
and circumnuclear regions are revealed.  The improvement in angular resolution
is approximately a factor of 3 to 4, with the resulting FWHM $\sim$3$\arcsec$
for W1, W2 and W3, and $\sim$6$\arcsec$ for W4.

\begin{figure*}[ht!]
\begin{center}
\includegraphics[width=16cm]{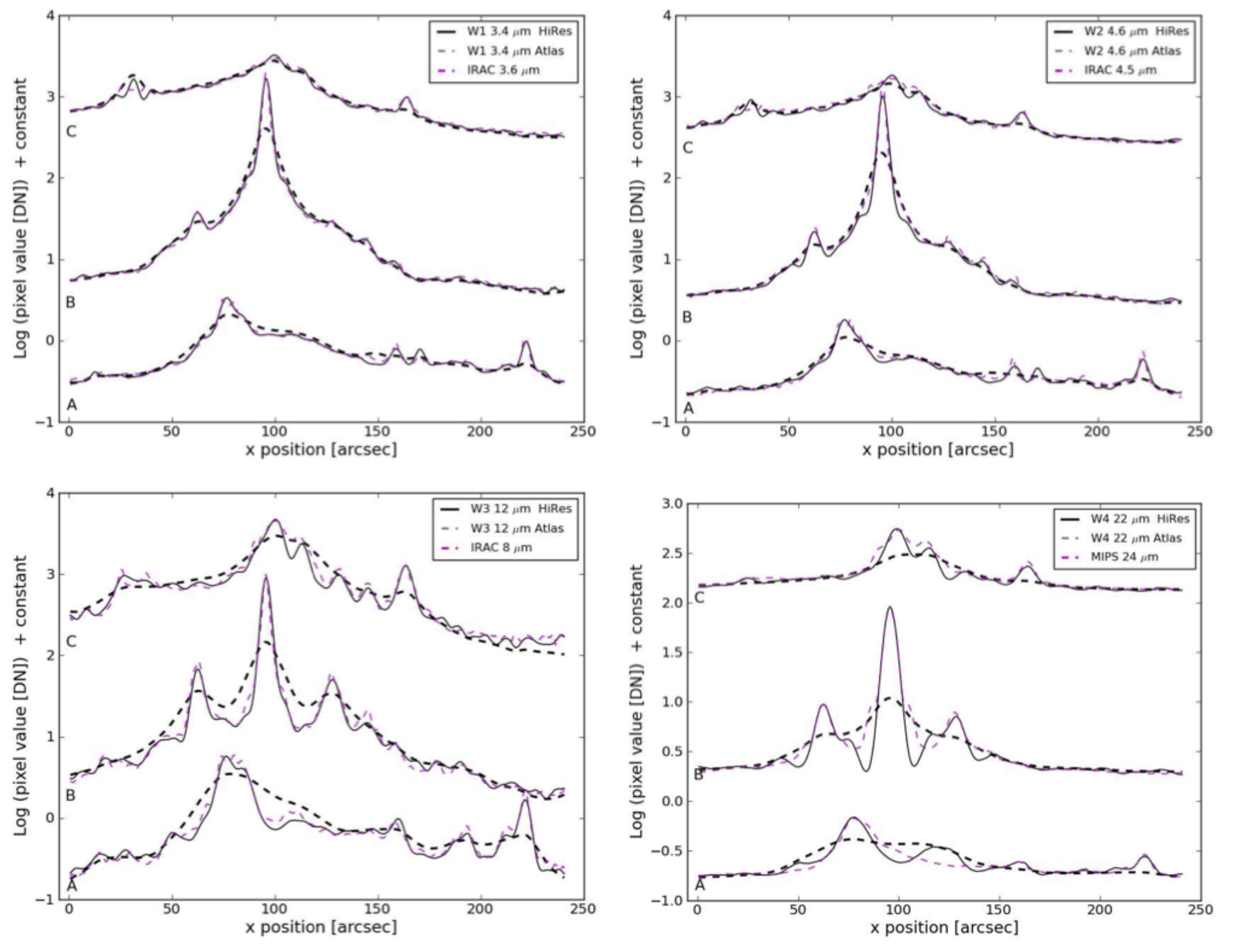}
\caption[line cuts]
{\small{1-D line profile comparison between the Atlas/nominal (dashed line), high-resolution (solid line)
and \Spitzer IRAC or MIPS24 (magenta dashed)
imaging of NGC\,1566.  Three profiles or line-cuts are shown, labelled A, B, C,
Fig. \ref{HiResBW} specifies the regions, with the center profile (B) slicing across
the nucleus and spiral arms.
}}
\label{line cuts}
\end{center}
\end{figure*}

Another way to view the
relative improvement in performance at large scales is through 1-D slices, cutting across the nucleus
and spiral arms north and south of the center (see Fig. \ref{line cuts}).   To gauge the
resolution fidelity of the WISE HiRes, also shown are the corresponding slices through IRAC and MIPS-24.
Roughly five different
spiral arms are crossed by the horizontal cuts.  Discrete structures
are more sharply delineated with HiRes and IRAC, while the overall, mean surface brightness
is conserved (see also Fig. \ref{HiRes3} below, showing the radial,
azimuthal average).

In particular, W3 is notably improved, the strong neutral-PAH emission arising within the spiral arms
and tracing star formation regions create the sawtooth pattern in the slices and
nucleus, inter-arm and spiral-arm features in the azimuthal average.   All of these
features are seen in the IRAC profiles, validating the reconstruction.
W4 shows an unavoidable consequence of the MCM-HiRes method: the formation of dark (lower intensity) troughs
around regions with bright, rapidly-varying emission on compact spatial scales.
Much lower amplitude troughs appear in the MIP-24 image, verifying that the W4 HiRes does
have this artifact feature.
For example, note the
cut through the bright nucleus (W4 panel of Fig. \ref{line cuts}), likewise seen in the azimuthal average
of the radial surface brightness, where the W4 panel of Fig. \ref{HiRes3} shows the circular ``ring" trough at a radius $\sim$15$\arcsec$.
At small scales, the flux is conserved within a circular area that encompasses the source and `ring'.
Although this ``ringing" phenomenon is minimized in the current HiRes method, it is strongest
in the W4 reconstructions due to
the relatively earlier onset of saturation that can modify source profiles compared to the native PSF.
The higher background in general also has an impact on the ringing that is introduced into the reconstruction.
Caution is required when interpreting the W4 light distribution in proximity to bright sources (stars, nuclei, etc);
nevertheless, as we show in Section 4.3, the overall W4 integrated flux is conserved and the reconstructions maintain science-level quality.

\smallskip
At yet larger scales, we investigate the relative performance by azimuthally averaging the surface brightness, comparing
the resultant radial profiles.  The local background was derived using an annulus with an inner radius of 6.7 arcmin and a width
of 0.3 arcmin (data reduction is discussed in detail in Paper II).  Fig. \ref{HiRes3} shows the radial profiles extending down to the local background level,
roughly 24.9, 24.5, 22.1 and 18.3 mag arcsec$^{-2}$ for W1, W2, W3 and W4 respectively.
The only significant difference between the Atlas and HiRes radial profiles are the better resolved spiral arms
and nuclear regions, notably for the star formation sensitive W3 and W4 bands.  At low surface brightnesses, the HiRes
is clearly conserving flux relative to the Atlas images.

\begin{figure*}[ht!]
\begin{center}
\includegraphics[width=17cm]{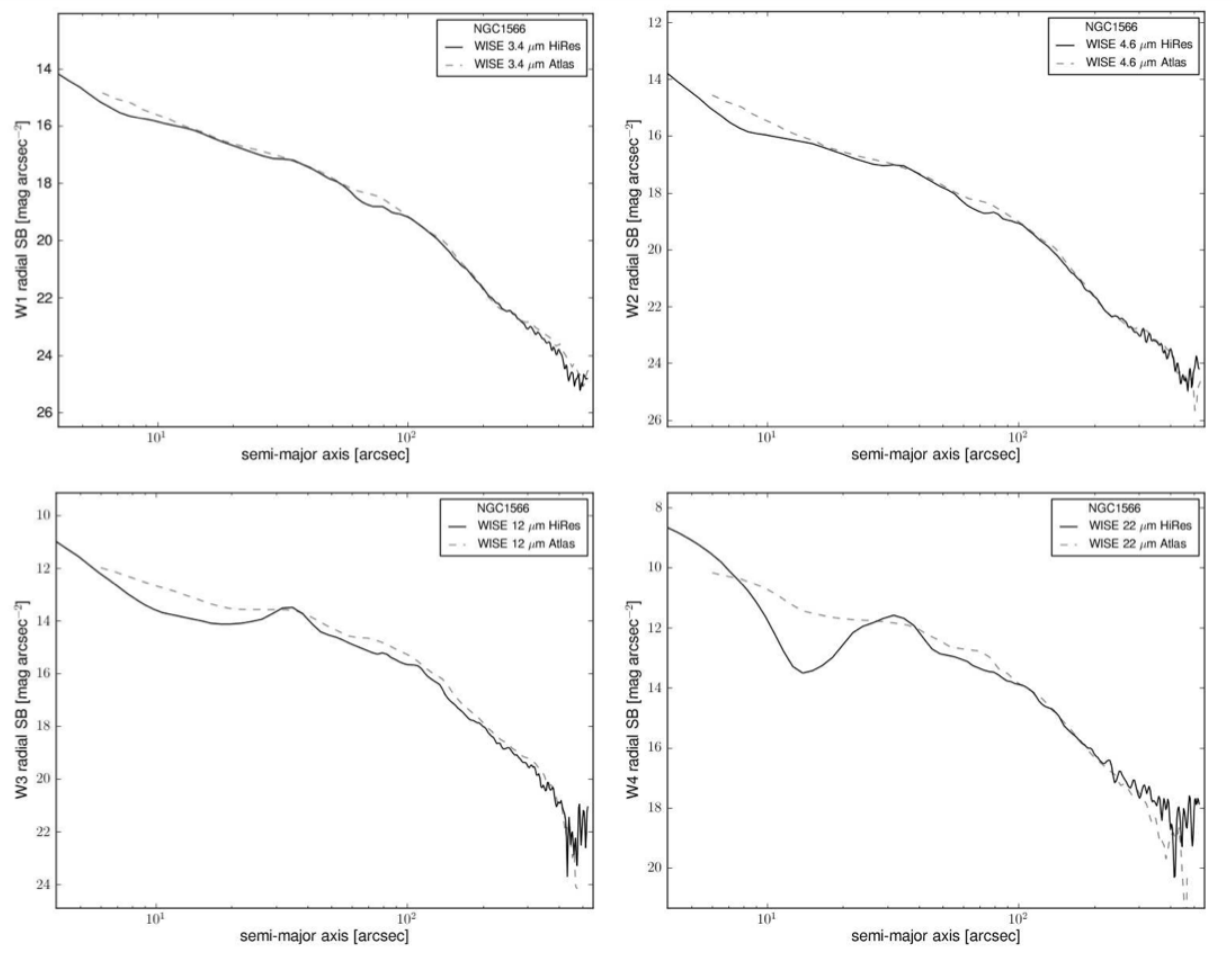}
\caption[HiRes3]
{\small{WISE azimuthally-averaged elliptical-radial profile of
NGC\,1566.  For comparison, the nominal resolution
and the high (HiRes) resolution are shown for
each band.
}}
\label{HiRes3}
\end{center}
\end{figure*}

\begin{figure*}[ht!]
\begin{center}
\includegraphics[width=17cm]{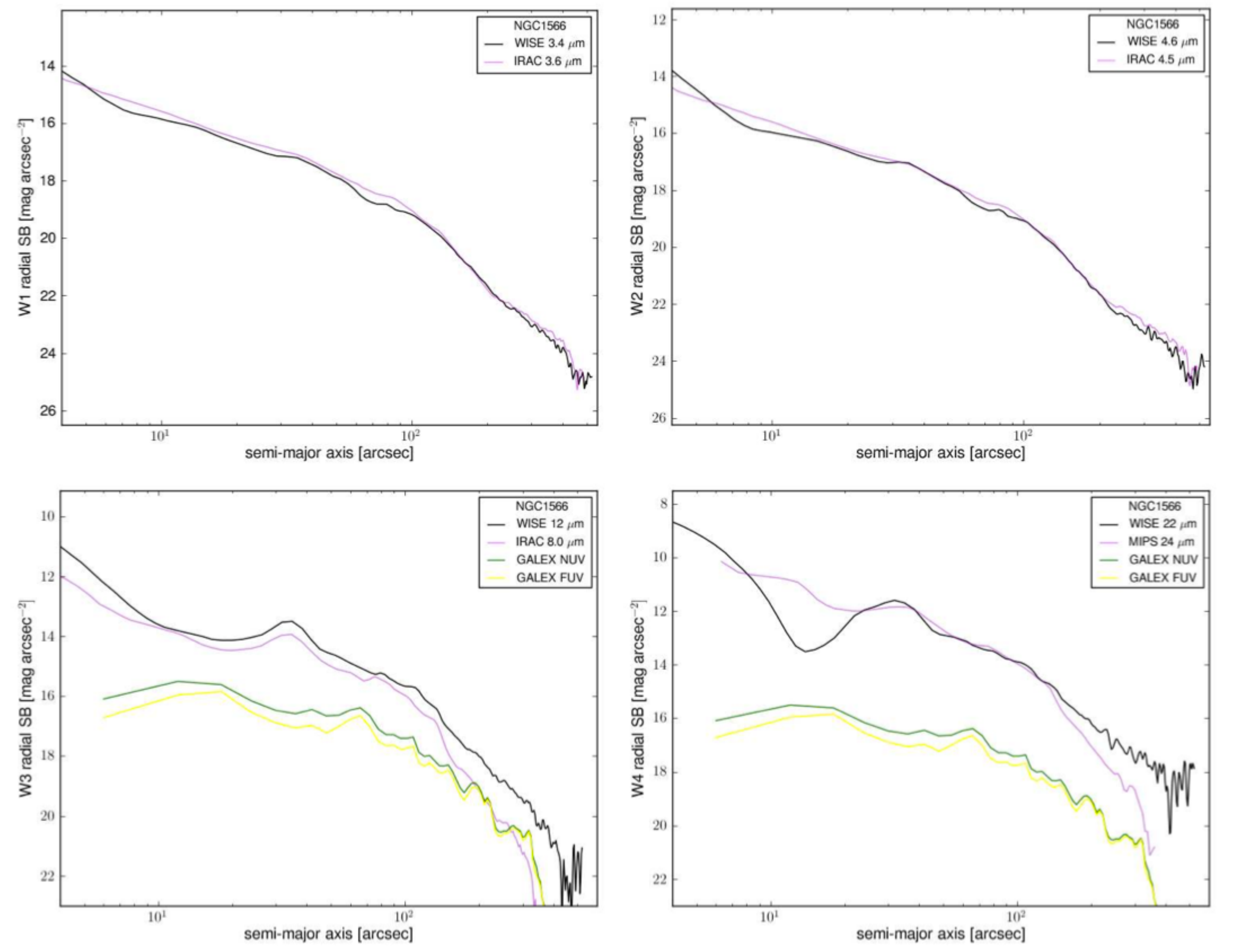}
\caption[HiRes5]
{\small{NGC\,1566 radial profile comparison between WISE HiRes, \Spitzer and GALEX.
The units are Vega mag arcsec$^{-2}$.  For the W3 and W4 panels, the
 GALEX AB magnitudes have been offset by $\sim$7 magnitudes to
fit within the Y-axis dynamic range.
 }}
\label{n1566radial}
\end{center}
\end{figure*}

We now compare the WISE HiRes radial surface brightness
profiles to those measured using \Spitzer and GALEX imaging.
Fig. \ref{n1566radial} presents the surface brightness for WISE (W1 and W2) and \Spitzer
(IRAC-1 and IRAC-2), and the addition of GALEX (NUV) to the W3+IRAC-4 profiles
and GALEX (FUV) to the W4+MIPS-24 profiles.    The WISE and \Spitzer magnitudes
are in standard Vega units, and the GALEX magnitudes are in AB units plus an offset of
$-$7 magnitudes to fit within the plot limits.

The first panel shows that W1 3.4 $\mu$m
and IRAC-1 3.6$\mu$m have very similar profiles, both easily reaching depths of
$\sim$24 mag arcsec$^{-2}$ (corresponding to 26.7 mag arcsec$^{-2}$ in AB).
The profiles are smooth, tracing the old stellar populations that form the spheroid
spatial distribution.
The second panel shows the W2 4.6 $\mu$m and IRAC-2 4.5$\mu$m
profiles, which are also nicely co-aligned.
The W2 and IRAC-2 depth reaches a similar limit
of $\sim$24 mag arcsec$^{-2}$ (corresponding to 27.3 mag arcsec$^{-2}$ in AB).
Both of these short-wavelength sets, W1/IRAC-1 and W2/IRAC-2,
have bandpasses that are similar enough that less than 5 to 10\% deviations are expected
due to spectral differences (detailed band-to-band analysis is presented in Paper II).
The third panel shows the W3 12 $\mu$m, IRAC-4 8 $\mu$m,
and GALEX NUV + FUV.  Both W3 and IRAC-4 have the same shape (radius $<$ 150 arcsec), but are offset slightly
due to band-to-band differences between WISE and \Spitzer:  we would expect
a flux ratio of 1.1 for late-type galaxies (see Paper II).  Also the IRAC-4 8 $\mu$m surface brightness is falling
rapidly beyond a radius of 200 arcsec, likely due increased noise and an asymmetric background gradient that
is created by scattered light and detector muxbleed from a nearby bright star
(see Fig. 1, also see Fig. 11 below).
The W3 depth reaches a limit
of 22 mag arcsec$^{-2}$ (corresponding to 27.2 mag arcsec$^{-2}$ in AB).
In the UV window, the profiles have considerably different shape;  notably, the absence of
UV emission in the central core (note the shallow profiles out to a radius of
20$\arcs$ radius) and the spiral arms shifted outward by 15 to 20$\arcs$
($\sim$0.9 Kpc), exhibiting a more pronounced, localized or compact, signal relative to the infrared.
The UV light then falls off steeper than the infrared light between 300 and 400$\arcs$, but then dramatically
extends well beyond ($>$ 500$\arcs$) the infrared signal, forming an additional set of arms that are invisible
to WISE and barely seen in IRAC-4  (see below, Fig \ref {3im}).
The last panel shows
the W4 22 $\mu$m, MIPS 24 $\mu$m,
and GALEX NUV + FUV.   W4 exhibits sharper profiles than MIPS-24 (note the dip
at 156$\arcs$) and exhibits higher surface brightness at larger radii ($>$200 arcsec),
MIPS-24 is also subject to a background gradient that is limiting the extraction at faint depths.
The W4 depth easily reaches a limit
of 18.5 mag arcsec$^{-2}$ (corresponding to 25.1 mag arcsec$^{-2}$ in AB).
Likewise with the W3 comparison to the UV, the W4 band appears shifted relative to
the GALEX bands, although it is not as distinct as 12 $\mu$m comparison.

\smallskip

The final set of comparisons to be made in this NGC\,1566 analysis is concerned with the
smallest scales.  To demonstrate the power of HiRes and its ability to tease out structure at
small scales, we zoom into the southeastern, inner spiral arm.
Fig \ref {3im} shows the star formation sensitive imaging of NGC\,1566:  WISE 12 12 $\mu$m HiRes
which is tracing the 11.3 $\mu$m PAH emission (predominently arising from
photodissociation regions or PDRs), \Spitzer-IRAC 8 $\mu$m, which is tracing
the 6.2 and 7.7 PAHs,  optical H$\alpha$, which traces recombination in HII regions,
and GALEX NUV (0.227 $\mu$m), which traces the high-energy stellar emission.   Both the H$\alpha$ and
the NUV are tracing the youngest, most massive populations, while the infrared arises from
obscured star formation from both massive and intermediate mass stars.

\begin{figure*}[ht!]
\begin{center}
\includegraphics[width=17cm]{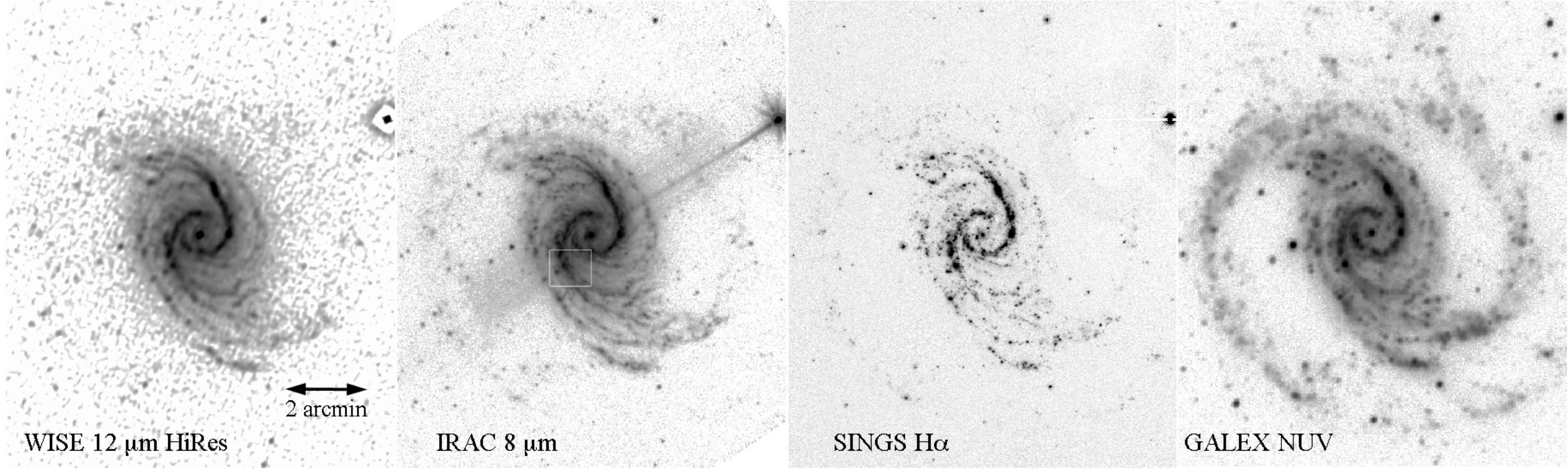}
\caption[3im]
{\small{Mapping the star formation in NGC\,1566 using infrared,
ultraviolet and H$\alpha$ tracers.  From left to right:
WISE 12 $\mu$m HiRes, \Spitzer-IRAC 8 $\mu$m, optical H$\alpha$ (0.656 $\mu$m), and
GALEX NUV (0.227 $\mu$m) imaging.  The small white box in the second
panel denotes the region that is highlighted in Fig. \ref{6grid}.
}}
\label{3im}
\end{center}
\end{figure*}

\begin{figure*}[ht!]
\begin{center}
\includegraphics[width=17cm]{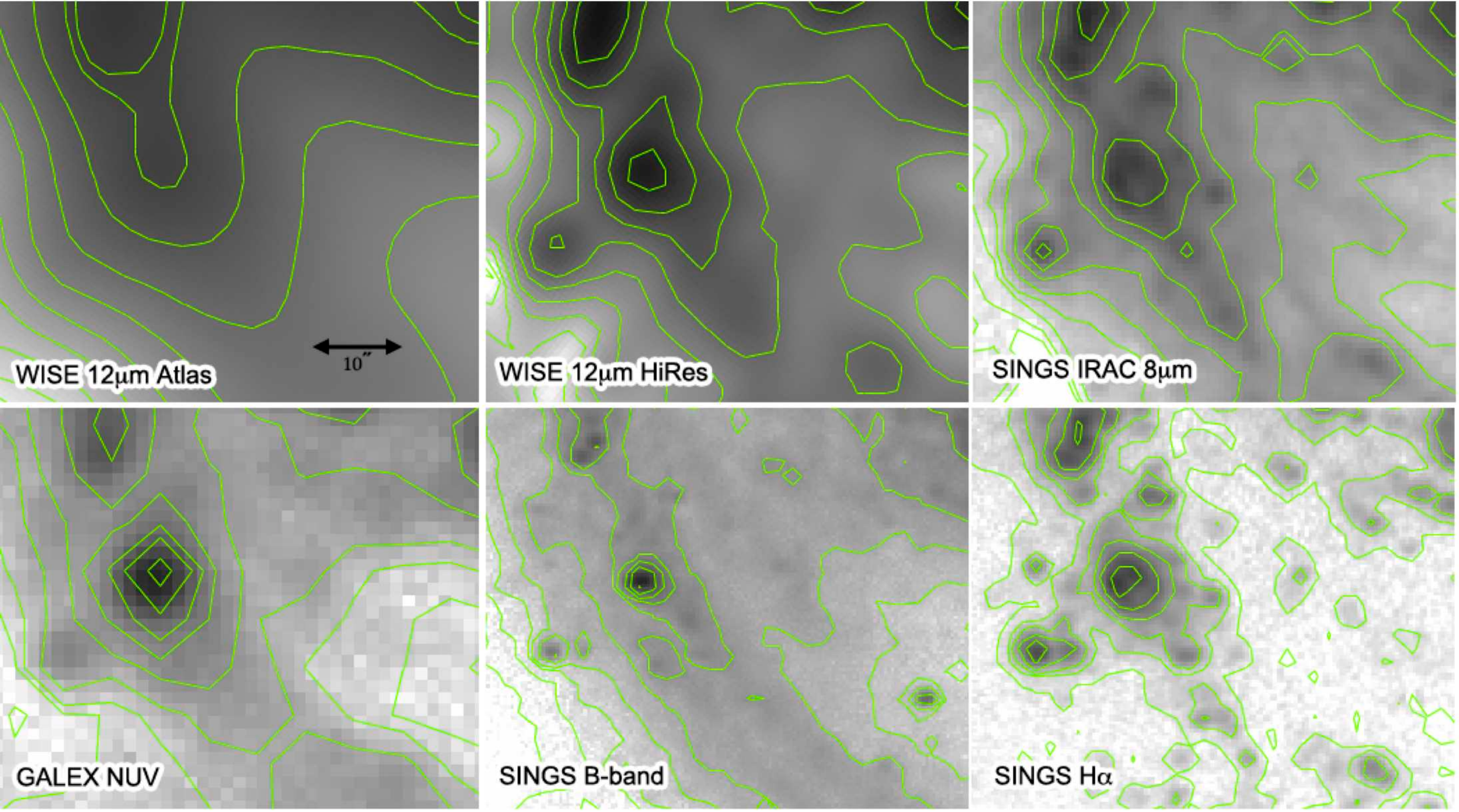}
\caption[6grid]
{\small{South-eastern spiral arm of NGC\,1566, centered on
RA = 04:20:04, DEC = -54:57:03 (see Fig. \ref{3im})..  The grey scale images show the star formation
sensitive bands of WISE 12 $\mu$m, \Spitzer-IRAC 8 $\mu$m,,
GALEX NUV (0.227 $\mu$m), optical B-band (0.44 $\mu$m) and optical H$\alpha$ (0.656 $\mu$m).
The green contours are used to enhance the star-formation structures captured
in the imaging.  The 10 $\arcs$ angular scale corresponds to a physical scale of 460 pc.
}}
\label{6grid}
\end{center}
\end{figure*}

The spiral arms and high surface brightness knots/filaments,
where the newly forming stars are primarly concentrated, are clearly delineated
for all four tracers.  Note that the sensitivity of the GALEX image, easily detecting the
outer spiral arm of this galaxy (yet invisible to WISE and nearly so to IRAC).
Zooming into the southeastern arm (as denoted in the 2nd panel), the
smallest scales that the imaging provides is shown in Fig. \ref{6grid}.

Six panels are shown in Fig \ref{6grid}: the upper panels represent the
obscured star formation, and the lower panels the unobscured star formation.
The WISE Atlas image is smooth and mostly featureless due to its poor angular
resolution, washing out the spiral arm and dense star formation complexes.
The situation is greatly improved after MCM super-resolution treatment,
the WISE image is now delineating the spiral arm into several components.
The overall and detailed morphology that the WISE HiRes image is revealing is validated and confirmed
with the \Spitzer-IRAC 8 $\mu$m image.  The knots, gradients and shape of the
region are reveal in WISE are easily seen in \Spitzer:  this confirms that the MCM process
is stable and reliable, recovering spatial information that is astrophysical and real.
Comparing to the unobscured, massive star formation tracers (lower panels), the
primary site of star formation (high surface brightness features) are co-spatial.
The H$\alpha$ image, with its superior angular resolution, resolves many of
these bright knots into smaller components (on physical scales the size of Giant
Molecular Clouds), but is only sensitive to the high surface brightness population
(HII regions).  Both the infrared and NUV trace more diffuse emission that is
associated with star formation from less massive (yet far more numerous) stars.

\smallskip

\begin{figure*}[ht!]
\begin{center}
\includegraphics[width=17cm]{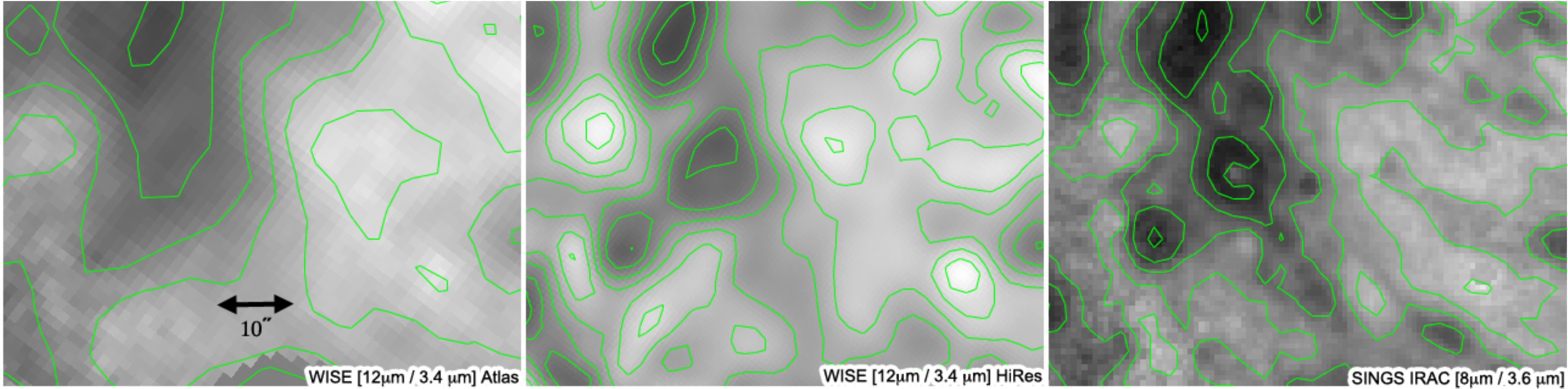}
\caption[sSFR]
{\small{Tracing the specific star formation rate (sSFR) in
the south-eastern spiral arm of NGC\,1566.
The proxy for the star formation rate is the mid-infrared (PAH)
emission, while the stellar mass is traced by the near-infrared emission.
Accordingly, we construct a (proportional) sSFR using the ratio of the WISE
12 to 3.4 $\mu$m imaging, and the IRAC
8 to 3.6 $\mu$m imaging.  See Fig. \ref{6grid}
for coordinate and scale information.
}}
\label{sSFR}
\end{center}
\end{figure*}

The nominal WISE Atlas images
are fully adequate to study the global characteristics of galaxies, but
 as Fig \ref{6grid} clearly demonstrates, the HiRes technique is crucial if
WISE is to be used to study the detailed anatomy of galaxies. As a final demonstration,
we now combine the power of the WISE bands to understand galaxy evolution.
The W1 (and W2) band is sensitive to the dominate mass component of (most) galaxies,
namely the evolved stellar population.  In contrast, the W3 (and W4) is sensitive to
the present-day star formation activity.   Combining the two creates a metric that
is used to assess the present to past star formation history.  This metric is also
known as the specific star formation rate, generally stated as the ratio of
the star formation rate (SFR) to the stellar mass (M$_*$).  For this demonstration,
we simply ratio the W3 image (which is proportional to the SFR, as detailed in Paper II),
to the W1 image (which is proportional to M$_*$ plus a color term, detailed in Paper II).
The result is shown in Fig. \ref{sSFR}, where we are comparing the WISE
sSFR with the equivalent \Spitzer sSFR (IRAC 8 to 3.6 $\mu$m ratio).
It is no surprise that the WISE Atlas result is mostly washed out, only painting
a gentle gradient across the field (i.e., the spiral arm itself).   The HiRes result,
however, reveals several concentrations of high sSFR (as confirmed by the
\Spitzer result).  These regions of enhanced sSFR are precisely where the
galaxy is creating new building blocks, where it is actively building its disk.
In contrast, the low sSFR regions, dominated by the inner-arm regions,
are currently quiescent and filled with older, evolved stars.  A metric such as the sSFR
demands the highest resolution possible since it is a ratio of two separate images,
and only WISE HiRes provides enough information to study the details star formation
evolution in galaxies.

As a final note to the analysis presented in this work,
NGC1566 is not an exception, the spatial resolution improvement from MCM-HiRES is realized for all of the galaxies in the sample presented in
Paper II, which assesses the photometric
performance of the reconstructed images in comparison to \Spitzer photometry.

\section{Summary}

In this paper we have presented a method by which to recover spatial information and
significantly improve the
angular resolution of WISE mid-infrared imaging, enabling detailed study of the internal
anatomy of galaxies.
Developed by the WISE Science Data Center, the Maximum Correlation
Method (MCM) yields improvements that are 3 to 4 times better than the
nominal WISE Atlas imaging resolution, and factors of 2 to 3 times improvement
from standard `drizzle' co-addition.    Using the nearby galaxy NGC\,1566 as a case study,
we
demonstrate how the
angular resolution of WISE may be accurately enhanced to achieve information on
physical scales
comparable to those of \Spitzer imaging.
This method will be used to construct the
WISE High Resolution Galaxy Atlas (WHRGA), consisting of several thousand
nearby galaxies.  A pilot study is now underway, the initial results
derived from a sample of 17 large, nearby galaxies
is presented in a
companion paper.

\newpage

\noindent Acknowledgements

This work is based [in part] on observations made with the {\it Spitzer} and research using the  NASA/IPAC Extragalactic Database (NED) and
IPAC Infrared Science Archive,
all are operated by JPL, Caltech under a contract with the National Aeronautics and Space Administration.
Support for this work was provided by NASA through an award issued by JPL/Caltech.
R.J.A. was supported by an appointment to the NASA Postdoctoral Program at the Jet Propulsion Laboratory, administered
by Oak Ridge Associated Universities through a contract with NASA.
This publication makes use of data products from the Wide-field Infrared Survey Explorer, which is a joint project of the University of California, Los Angeles, and the Jet Propulsion Laboratory/California Institute of Technology, funded by the National Aeronautics and Space Administration.

\pagebreak
\section{References}

\noindent Aumann, H. H., Fowler, J. W., \& Melnyk, M. 1990, AJ, 99, 1674\\
\noindent Cutri, R., et al., 2012, WISE Explanatory Supplement\\
\noindent Fowler, J. W., \& Aumann, H. H. 1994, in Science with High-Resolution Far-Infrared Data, ed. S. Terebey \& J. Mazzarella (JPL Publication 94-5), 1.\\
\noindent Jarrett, T.J., et al., 2011, ApJ, 735, 112\\
\noindent Jarrett, T.J., et al., 2012, AJ (Paper II)\\
\noindent Kennicutt, R. et al. 2003 PASP, 115, 928\\
\noindent Martin, D. C., et al. 2005, ApJ, 619, L1\\
\noindent Masci, F.J., \& Fowler, J.W.,  in Proceedings of Astronomical Data Analysis Software and Systems XVIII, Québec City, ASP Conference Series, Edited by D. Bohlender, P. Dowler, and D. Durand, Vol. 411, 2009, p.67\\
\noindent Morrissey, P., et al. 2005, ApJ, 619, L7\\
\noindent Morrissey, P., 2007, ApJS, 173, 682\\
\noindent Murakami, H., et al., 2007, PASJ, 59, S369.\\
\noindent Sheth, K., et al, 2010, PASP, 122, 1397\\
\noindent Tully, R.B. 1988, NEARBY GALAXY CATALOG\\
\noindent Willick, J.A., et al., 1997, ApJS, 109, 333\\
\noindent Wright, E. et al., 2010, AJ, 140, 1868\\


\begin{table}[ht!]
\caption{Point Source Profile Widths}
\begin{center}
\begin{tabular}{r r r r r r r r r r r}

\hline
\hline
\\[0.25pt]

method   & W1 3.4 $\mu$m  & W2 4.6 $\mu$m  & W3 12$\mu$m & W4 22 $\mu$m   \\
           & FWHM [arcsec] &   FWHM [arcsec]  & FWHM [arcsec]  & FWHM [arcsec]  \\

\hline
\\[0.25pt]

Atlas     &  8.1 &  8.8&   11.1&   17.5\\
drizzle  & 5.9 & 6.5 &  7.0 & 12.4 \\ 
MCM HiRes & 2.6 & 3.0 & 3.5 & 5.5 \\

\hline
\end{tabular}
\end{center}
\tablecomments{Atlas co-added images are the standard public release product of
WISE.  Variable-Pixel Linear Reconstruction, or ``drizzle" , and the
MCM-HiRes method are part of the WISE High Resolution Galaxy Atlas.
}


\end{table}

\end{document}